\documentclass[conference]{IEEEtran}
\usepackage{amsmath,amssymb,amscd,cite}
\usepackage[alwaysadjust]{paralist}
\usepackage{balance}
\newtheorem{thm}{Theorem}[section]
\newtheorem{prop}[thm]{Proposition}
\newtheorem{lem}[thm]{Lemma}
\newtheorem{cor}[thm]{Corollary}

\newcommand{\pf}{{\bf Proof. \ }}
\newcommand{\qed}{\hfill $\Box$ \\}
\font\msbm=msbm10 at 12pt
\newcommand{\Z}{\mbox{\msbm Z}}

\newtheorem{ex}[thm]{Example}

\newcommand{\ord}{ord}

\begin{document}
\title{Self-dual Repeated Root Cyclic and Negacyclic Codes over Finite Fields}
\author{
\authorblockN{K. Guenda}
\authorblockA{Faculty of Mathematics USTHB\\
University of Sciences and Technology of Algiers\\
B.P. 32 El Alia, Bab Ezzouar, Algiers, Algeria\\
Email: kguenda@gmail.com} \and
\authorblockN{T. A. Gulliver}
\authorblockA{Dept. of Electrical and Computer Engineering\\
University of Victoria\\
P.O. Box 3055, STN CSC, Victoria, BC Canada V8W 3P6\\
Email: agullive@ece.uvic.ca}}
\maketitle
\begin{abstract}
In this paper we investigate repeated root cyclic and negacyclic
codes of length $p^rm$ over $\mathbb{F}_{p^s}$ with $(m,p)=1$. In
the case $p$ odd, we give necessary and sufficient conditions on the
existence of negacyclic self-dual codes.
When $m=2m'$ with $m'$ odd, we characterize the codes in terms of their
generator polynomials. This provides simple conditions on the
existence of self-dual negacyclic codes, and generalizes the
results of Dinh~\cite{dinh}. We also answer an open problem
concerning the number of self-dual cyclic codes given by Jia et al.~\cite{jia}.
\end{abstract}

\section{Introduction}

Let $p$ be a prime number and $\mathbb{F}_{p^s}$ the finite field
with $p^s$ elements. An $[n,k]$ linear code $C$ over
$\mathbb{F}_{p^s}$ is a $k$-dimensional subspace of
$\mathbb{F}_{p^s}^n$. A linear code $C$ over $\mathbb{F}_{p^s}^n$ is
said to be constacyclic if it is an ideal of the quotient ring
$R_n=\mathbb{F}_{p^s}[x]/\langle x^n-a \rangle$. When $a =1$ the
code is called cyclic, and when $a=-1$ the code is called negacyclic.
The Euclidean dual code $C^{\bot}$ of $C$ is defined as
$C^{\bot}=\{\mathsf{x} \in \mathbb{F}_q^n : \sum_{i=1}^{n}x_iy_i=0
\, \forall \, \mathsf{y} \in C\}$. An interesting class of codes is
the so-called self-dual codes. A code is called Euclidean self-dual
if it satisfies $C=C^{\bot}$. Note that the dual of a cyclic
(respectively negacyclic) code is a cyclic (respectively negacyclic)
code.

Cyclic codes are interesting from both theoretical and practical
perspectives. For example, they can easily be encoded, and decoding
algorithms exist in many cases. When $(n,p)=1$, these codes are
called simple root codes, otherwise they
are called repeated root codes. Castagnoli et
al. and van Lint~\cite{castagnoli,vanlint} studied repeated root cyclic codes.
They proved that these codes have a concatenated structure and are
not asymptotically better than simple root codes.
Negacyclic codes were introduced by
Berlekamp~\cite{berl1}. Simple root self-dual negacyclic codes were
studied by Blackford~\cite{blackford} and Guenda~\cite{guenda011}.
The algebraic structure of repeated root constacyclic codes of
length $2p^r$ over $\mathbb{F}_{p^s}$ as well as the self-duality of
such codes has also been investigated by Dinh~\cite{dinh}.
Conditions on the existence of cyclic self-dual codes of length
$2^rm$ over $\mathbb{F}_{2^s}$ were studied independently by Kai and
Zhu~\cite{xai} and Jia et al.~\cite{jia}. Jia et al. also determined
the existence and the number of cyclic self-dual codes for $q=2^m$.

In this paper, we investigate repeated root cyclic and negacyclic
codes of length $p^rm$ over $\mathbb{F}_{p^s}$ with $(m,p)=1$. When
$p$ is odd, we give necessary and sufficient conditions on
the existence of negacyclic self-dual codes.
When $m=2m', m'$ odd,
we determine explicitly the generator polynomials using ring
isomorphisms. This provides simple conditions on the existence of
negacyclic self-dual codes. We also answer an open problem
concerning the number of self-dual cyclic codes given by Jia et
al. \cite{jia}.

\section{Self-dual Negacyclic Codes of Length $mp^r$ over $\mathbb{F}_{p^s}$}

Throughout this section, $p$ is an odd prime number and $n=mp^r$,
with $m$ an integer (odd or even) such that $(m,p)=1$. This section
provides conditions on the existence of self-dual negacyclic codes
of length $n=mp^r$ over $\mathbb{F}_{p^s}$. It is well known that negacyclic
codes over $\mathbb{F}_{p^s}$ are principal ideals generated by the factors
of $x^{mp^r}+1$. Since $\mathbb{F}_{p^s}$ has characteristic $p$, the
polynomial $x^{mp^r}+1$ can be factored as
\begin{equation}
\label{eq:factor2} x^{mp^r}+1=(x^{m}+1)^{p^r}.
\end{equation}
The polynomial $x^{m}+1$ is a monic square free polynomial, hence
from~\cite[Proposition 2.7]{permounth} it factors uniquely as a
product of pairwise coprime monic irreducible polynomials $f_1(x),
\ldots, f_l(x)$. Thus from~(\ref{eq:factor2}) we obtain the
following factorization of $x^{mp^r}+1$
\begin{equation}
\label{eq:factor3} x^{mp^r}+1={f_1(x)}^{p^r}\ldots {f_l(x)}^{p^r}.
\end{equation}
A negacyclic code of length $n=mp^r$ over $\mathbb{F}_{p^s}$ is then
generated by a polynomial of the form
\begin{equation}
\label{eq:gen}
 A(x)=\prod {f_i}^{k_i},
 \end{equation}
where $f_i(x), i \le l$, are the polynomials given in
(\ref{eq:factor3}) and $0 \le k_i \le p^r$.

For a polynomial $f(x)=a_0+a_1 x\ldots+ a_rx^r$, with $a_0 \neq 0$
and degree $r$ (hence $a_r\neq 0$), the reciprocal of $f$ is the
polynomial denoted by $f^*$ and defined as
\begin{equation}
\label{eq:5} f^*(x)=x^rf(x^{-1})=a_r+a_{r-1}x+\ldots + a_0x^r.
\end{equation}
If a polynomial $f$ is equal to its reciprocal, then $f$ is called
self-reciprocal. We can easily verify the following equalities
\begin{equation}
\label{eq:prop}
 (f^*)^*=f \text{ and } (fg)^*=f^*g^*.
\end{equation}

It is well known (see \cite[Proposition 2.4]{dinh} or~\cite[Theorem
4.4.9]{huffman03}), that the dual of the negacyclic code generated
by $A(x)$ is the negacyclic code generated by $B^*(x)$ where
\begin{equation}
\label{eq:dual} B(x)=\frac{x^n+1}{A(x)}.
\end{equation}
Hence we have the following lemma.
\begin{lem}
\label{lem:dual} A negacyclic code $C$ of length $n$ generated by a
polynomial $A(x)$ is self-dual if and only if
\[
A(x)=B^*(x).
\]
\end{lem}
Denote the factors $f_i$ in the factorization of $x^m+1$ which are
self-reciprocal by $g_1, \ldots g_s$, and the remaining $f_j$
grouped in pairs by $h_1,h_1^*,\ldots, h_t,h_t^*$. Hence $l=s+2t$,
and the factorization given in~(\ref{eq:factor3}) becomes
\begin{equation}
\begin{array}{ccl}
\label{eq:ling} x^n+1&=&(x^m+1)^{p^r}=g^{p^r}_1(x)\ldots
g^{p^r}_s(x)\\
&&\times h^{p^r}_1(x)h^{* p^r}_1(x)\ldots h^{p^r}_t(x)h^{*
p^r}_t(x).
\end{array}
\end{equation}
\begin{thm}
\label{th:exist} There exists a self-dual negacyclic code of length
$mp^r$ over $\mathbb{F}_{p^s}$ if and only if there is no $g_i$
(self-reciprocal polynomial) in the factorization of $x^{mp^r}+1$
given in~(\ref{eq:ling}). Furthermore, a self-dual negacyclic code
$C$ is generated by a polynomial of the following form
\begin{equation}
h^{b_1}_1(x)h^{* p^r- b_1}_1(x)\ldots h^{b_t}_t(x)h^{*
p^r-b_t}_t(x).
\end{equation}
\end{thm}
\pf Assume there exists a negacyclic self-dual code $C$ of length
$n=mp^r$ over $\mathbb{F}_{p^s}$. Hence from (\ref{eq:gen}) the code $C$ is
generated by $A(x)= \prod {f_i}^{k_i},$ where the $f_i$ are factors
of $x^m+1$. From~(\ref{eq:ling}), we can write
\[
A(x)=g^{a_1}_1(x)\ldots g^{a_s}_s(x)h^{b_1}_1(x)h^{* c_1}_1(x)\ldots
h^{b_t}_t(x)h^{* c_t}_t(x),
\]
where $0 \le a_i \le p^s$ for $ 1 \le i \le s$, and $0 \le b_j \le
p^s$ and $0 \le c_j \le p^s$ for  $ 1 \le j \le t$.
Let $B(x)=\frac{x^{mp^s}+1 }{A(x)}$, and substituting $A(x)$ above gives
\[
\begin{array}{ccl}
B(x)&=&g^{p^r-a_1}_1(x)\ldots g^{p^r-a_s}_s(x)\\
&&\times h^{p^r-b_1}_1(x)h^{* p^r-c_1}_1(x)\ldots
h^{p^r-b_t}_t(x)h^{* p^r-c_t}_t(x).
\end{array}
\]
Using (\ref{eq:prop}) repeatedly in the factorization of $H(x)$, we
obtain
\[
\begin{array}{ccl}
B^*(x)&=&g^{p^r-a_1}_1(x)\ldots g^{p^r-a_s}_s(x)\\
&&\times h^{*p^r-b_1}_1(x)h^{ p^r-c_1}_1(x)\ldots
h^{*p^r-b_t}_t(x)h^{ p^r- c_t}_t(x).
\end{array}
\]

Since $C$ is self-dual, from~Lemma~\ref{lem:dual} we have that
$A(x)=B^*(x)$, and then by equating factors of $A(x)$ and $B^*(x)$,
the powers of these factors must satisfy $a_i =
p^s-a_i$ for $1\le i \le s$, and $b_j= p^s- b_j$ for $1 \le j \le
t$. Equivalently, $p^s =2a_i$ for $1\le i \le s$, and $c_j=
  p^s- b_j$ for $1 \le j \le t$. Since $p$ is odd, the last equalities
are possible if and only if there is no $g_i$ in the factorization
of $x^{mp^s}+1$ and $c_j= p^s- b_j,$ for $1 \le j \le t$, i.e, $s=0$
in (\ref{eq:ling}) and $c_j=p^s- b_j,$ for $1 \le j \le t$.
Hence a negacyclic self-dual code is generated by
\[
h^{b_1}_1(x)h^{* p^r-b_1}_1(x)\ldots h^{b_t}_t(x)h^{* p^r-b_t}_t(x).
\]
\qed

\begin{lem}
\label{Prop: 3.1} Let $p^s$ be an odd prime. Then the following
holds
\begin{enumerate}
\item[(i)] If $p\equiv 1 \mod 4$, $s$ any integer or $p \equiv 3 \mod 4$
and $s$ even, then $x^2+1=0$ has a solution $\gamma \in
\mathbb{F}_{p^s}$.
\item[(ii)] If $p\equiv 3 \mod 4$, $s$ odd, then $x^2+1$ is irreducible in $\mathbb{F}_{p^s}$.
\end{enumerate}
\end{lem}
\pf Since $p \equiv 1 \pmod 4$, $-1$ is a quadratic residue in
$\mathbb{F}_p \subset \mathbb{F}_{p^s}$~\cite[Lemma
6.2.4]{huffman03}. Thus there exists $\gamma \in \mathbb{F}_{p^s}$
such that $\gamma^2=-1$. If $p \equiv 3 \pmod 4$, then $p^2 \equiv 1
\pmod 4$, so that $-1$ is a quadratic residue in $\mathbb{F}_{p^2}
\subset \mathbb{F}_{p^{s}}$.
The proof of (ii) is in~\cite[Proposition 3.1 (ii)]{dinh}. \qed

\section{Negacyclic Codes of Length $2mp^r$ over $\mathbb{F}_{p^s}$}

In this section, we consider the structure of negacyclic codes over
$\mathbb{F}_{p^s}$ of length $2mp^r$. We begin with the following lemma.
When $(m,p)=1$, $m$ an odd integer, Dinh and
L\'{o}pez-Permouth~\cite[Proposition 5.1]{permounth} proved that
negacyclic codes of length $m$ are isomorphic to cyclic codes.
Batoul et al.~\cite{aicha2011} proved that under some conditions,
there also exists an isomorphism between constacyclic codes and
cyclic codes of length $m$. In the following lemma, we prove that
there is an isomorphism between cyclic codes and some constacyclic
codes with conditions different from those in
\cite{permounth,aicha2011}.
\begin{lem}
\label{lem:equiv} Let $p^s$ be an odd prime power such that $p\equiv
1 \mod 4$, $s$ any integer or $p \equiv 3 \mod 4$ and $s$ even. Then
there is a ring isomorphism between the ring $\frac{
\mathbb{F}_{p^s}[x]}{x^m-1}$
 and the ring
$\frac{\mathbb{F}_{p^s}[x]}{x^m-\gamma }$ given by
\[
\mu (f(x))=\left\{
\begin{array}{ll}
 f( \gamma x) &  \text{ if } m \equiv 3 \mod  4,\\
 f(- \gamma x)& \text{ if } m \equiv 1 \mod 4. \\
\end{array}
\right.
\]
Furthermore, there is a ring isomorphism between the ring
$\frac{\mathbb{F}_{p^s}[x]}{x^m-1}$ and the ring
$\frac{\mathbb{F}_{p^s}[x]}{x^m+\gamma}$ given by
\end{lem}
\[
\mu (f(x))=\left \{
\begin{array}{ll}
 f (-\gamma x)& \text{ if } m \equiv 3 \mod 4 , \\
 f (\gamma x) & \text{ if }  m \equiv 1 \mod 4. \\
\end{array}
\right.
\]
\pf From the assumptions on $p$ and $s$ in~Lemma~\ref{Prop: 3.1},
there exists a solution $\gamma$ to $x^2+1=0$. We only prove the
ring isomorphism between the ring $\frac{\mathbb{F}_{p^s}[x]}{x^m-1}$ and
the ring $\frac{\mathbb{F}_{p^s}[x]}{x^m-\gamma}$. The other isomorphism can
easily be obtained in a similar manner. Since $\gamma^2=-1$, we have
that $\gamma^m= \gamma$ if $m \equiv 1 \mod 4$, and $\gamma^m=
-\gamma$ if $m \equiv 3 \mod 4$. Assume that $m \equiv 3 \mod 4$, so
that $\mu f(x)= f(\gamma x)$ for  $ f(x) \in \mathbb{F}_{p^s}[x]$.
It is obvious that $\mu$ is a ring homomorphism, hence we only need to
prove that $\mu$ is a one-to-one map. For this, let $f(x)$ and
$g(x)$ be polynomials in $\mathbb{F}_{p^s}[x]$ such that
\[
f(x)\equiv g(x) \pmod {x^m-1}.
\]
This is equivalent to the existence of $h(x)\in \mathbb{F}_{p^s}[x]$ such
that $f(x)-g(x)=h(x)(x^m-1)$, and this equality is true if and only
if $f(\gamma x)-g(\gamma x)=h(\gamma x)((\gamma x)^m-1)$ is true.
The assumption on $m$ gives that $\gamma^m=-\gamma$. Then we have
$f(\gamma x)-g(\gamma x)=-\gamma h(\gamma x)(x^m-\gamma)$. This
equality is equivalent to $f(\gamma x)-g(\gamma x) \equiv 0 \mod
x^n-\gamma$. This means that for $f$ and $g$ in $\mathbb{F}_{p^s}[x]/\langle
x^m-1\rangle$, we have $f(x)= g(x)$ if and only if $\mu(f(x))=
\mu(g(x))$. Hence it follows that $\mu$ is an isomorphism. A similar
argument holds with $m \equiv 1 \mod 4$ for $\mu (f(x))=f(\ -\gamma
x)$.
 \qed
\begin{thm}
\label{thm:prod} Let  $p^s$ be an odd prime power such that $p\equiv
1 \mod 4$, $s$ any integer or $p \equiv 3 \mod 4$ and $s$ even, and
$n=2mp^r$ be an oddly even integer with $(m,p)=1$. Then a negacyclic
code of length $n$ over $\mathbb{F}_{p^s}$ is a principal ideal of
$\mathbb{F}_{p^s}[x]/\langle x^n+1 \rangle$ generated by a polynomial of the
following form
\[\prod_{i \in I} f_i^{t_i}(\gamma x) \prod_{j \in
J} f_j^{t_j}(-\gamma x),
\]
where $f_i(x),$ $f_j(x)$ are monic irreducible factors of $x^m-1$,
and $0 \le t_i,t_j \le p^s.$
\end{thm}
\pf  It suffices to find the factors of $x^{2mp^r}+1$.
From~Lemma~\ref{Prop: 3.1}, $x^2+1=0$ has a solution $\gamma \in
\mathbb{F}_{p^s}$, so $x^{2mp^r}+1$ can be decomposed as
$(x^{2m}+1)^{p^r}=(x^{m}+\gamma)^{p^r}(x^{m}-\gamma)^{p^r}$. The
result then follows from the isomorphisms given
in~Lemma~\ref{lem:equiv}. \qed
\begin{ex}
\label{ex:dinda} In the case $p\equiv 1 \mod 4$, $s$ any integer or
$p \equiv 3 \mod 4$ and $s$ even, $n=2p^r$, (i.e. $m=1$), there is
a unique factor of $x-1$ which is $f(x)=x-1$. Hence from
Theorem~\ref{thm:prod}, negacyclic codes of length $2p^r$ over
$\mathbb{F}_{p^s}$ are generated by
\begin{equation}
\label{eq:dinda}
 C =\langle (x-\gamma)^i(x+\gamma^j) \rangle, \text{ where }0 \le
i,j \le p^r.
\end{equation}
The result given in (\ref{eq:dinda}) was also proven
in~\cite[Theorem 3.2]{dinh}.
\end{ex}

\subsection{Self-dual Negacyclic Codes of Length $2mp^r$}

The purpose of this section is to provide conditions on the
existence of self-dual codes. This is done considering only the
length and characteristic. This gives equivalent conditions to those
in Theorem~\ref{th:exist} which are much simpler to verify. We first
present an example.
\begin{ex}
\label{ex:dinda2} For $m=1$, we have the following.
\begin{enumerate}
\item[(i)] If $p\equiv 1 \mod 4$, $s$ any integer or $p \equiv 3 \mod 4$
and $s$ even, then from~Theorem~\ref{th:exist} there exist self-dual
codes of length $2p^s$ over $\mathbb{F}_{p^s}$ if and only if none of the
irreducible factors of $x^2+1$ are self-reciprocal. From
Lemma~\ref{Prop: 3.1}, there is a solution $\gamma$ of $x^2+1=0$ in
$\mathbb{F}_{p^s}$. Hence the irreducible factors of $x^2+1$ are $x- \gamma$
and $x+\gamma$. Neither of these polynomials can be self-reciprocal,
as we have $(x- \gamma)^*=-\gamma(x+\gamma)$ and
$(x+\gamma)^*=\gamma(x-\gamma)$. Hence by Theorem~\ref{th:exist}
there exist negacyclic self-dual codes of the following form
\[\langle (x-\gamma)^i(x+\gamma)^{p^r-i}, \text{ where }0 \le i \le p^r. \]
\item[(ii)] If $p\equiv 3 \mod 4$, $s$ odd, then from Lemma~\ref{Prop: 3.1}
$x^2+1$ is irreducible in $\mathbb{F}_{p^s}$. Furthermore, we have
$(x^2+1)^*= x^2+1$. Hence by Theorem~\ref{th:exist} there are no
self-dual negacyclic codes in this case.
\end{enumerate}
\end{ex}
The results in Example~\ref{ex:dinda2} are also given
in~\cite[Corollary 3.3]{dinh}.

We now require the following Lemma.
\begin{lem}
\label{lem:rev1} Let $m$ be an odd integer and $Cl_m(i)$ the $p^s$
cyclotomic class of $i$ modulo $m$. The polynomial $f_i(x)$ is the
minimal polynomial associated with $Cl_m(i)$, hence we have
$Cl_m(i)=Cl_m(-i)$ if and only if $f_i(x)=f_i^*(x)$.
\end{lem}
\pf Let $\alpha$ be an $m$th primitive root of unity. The elements
of $Cl(i)$ are such that $\alpha^i$ is a root of a monic irreducible
polynomial $f_i(x)=a_0+a_1x+\ldots+x^r$. Hence $f_i^*(x)=x^{\deg
f_i}f_i(x^{-1})$ has $\alpha^{-i}$ as a root. Therefore
$Cl_m(i)=Cl_m(-i)$ if and only if the polynomials $f_i(x)$ and
$f_i^*(x)$ are monic with the same degree and the same roots, and
hence are equal.\qed
\begin{lem}
\label{lem:rev2} Let $m$ be an odd integer and $p$ a prime number.
Then $\ord_m(p^s)$ is even if and only if there exists a cyclotomic
class $Cl_m(i)$ which satisfies $Cl_m(i)=Cl_m(-i)$.
\end{lem}
\pf  Assume that $\ord_m(p^s)$ is even. We start with the case where
$m=q^{\alpha}$ is a prime power. We first prove the following
implication
\[
\ord_{q^{\alpha}}(p^r) \text{ is even }\Rightarrow\ord_q(p^r) \text{
is even} .
\]
Assume that $\ord_{q^{\alpha}}(p^r)$ is even and $\ord_q(p^r)$ is
odd. Then there exists odd $i>0$ such that $p^{ri}\equiv 1 \mod q
\Leftrightarrow p^{ri}=1+kq$. Hence $p^{riq^{\alpha
-1}}=(1+kq)^{q^{\alpha- 1}} \equiv 1\mod q^{\alpha}$, because $(1+
kq)^{q^{\alpha -1}} \equiv 1+kq^{\alpha}\mod q^{(\alpha+1)}$ (the
proof of the last equality can  be found in
\cite[Lemma~3.30]{demazure}). Therefore we have that
\begin{eqnarray}
\label{fermat} p^ {riq^{\alpha -1}} \equiv 1\mod q^{\alpha}.
\end{eqnarray}
If both $i$ and ${q^{\alpha -1}}$ are odd, then $\ord_{q^{\alpha}}(p^r)$ is odd,
which is absurd. Then it must be that $\ord_q(p^r)$ is even, so
there exists some integer $j$ such that $0<j< \ord_q (p^r)$ and
$p^j\equiv -1\mod q$.  Therefore we have $p^{rjq^{\alpha -1}}\equiv
-1\mod q^{\alpha}$, which gives that $Cl_{q^{\alpha}}(1)=
Cl_{q^{\alpha}}(-1)$. Then for all $i$ in the cyclotomic classes we
have $Cl_{q^{\alpha}}(i)=Cl_{q^{\alpha}}(-i)$. Assume now that
$m=p_1p_2$ such that $(p_1,p_2)=1$ and $\ord_m(p^s)$ is even. Since
$m=p_1p_2$, we have that
$\ord_m(p^s)=\mbox{lcm}(\ord_{p_1}(p^s),\ord_{p_2}(p^s))$ is even,
and hence either $\ord_{p_1}(p^s)$ or $\ord_{p_2}(p^s)$ is even.
Assume that $\ord_{p_1}(p^s)$ is even, then there exists $1\le k \le
\ord_{p_1}(p^s)$ such that $(p^s)^k\equiv -1 \mod p_1$. Therefore
$(p^s)^k(m-p_2)\equiv -(m-p_2) \mod m$, with $k \le \ord
_{p_1}(p^s)$, and hence $Cl_m(m-p_2)=Cl_m(-(m-p_2))$. The same
result is obtained for $m=p_1^{\alpha_1}p_2^{\alpha_2}$. Conversely,
assume there exists a class for which $Cl_m(i)=Cl_m(-i)$. Then the
elements of $Cl_m(i)$ are $\pm i q^j$ for some $j$, so
$Cl(i)$ contains an even number of elements. On the other hand the
size of each $q$ cyclotomic class is a divisor of
$\ord_m(p^s)$~\cite[Theorem 4.1.4]{huffman03}. This gives that
$\ord_m(p^s)$ is even. \qed

\begin{thm}
\label{th:self} Let  $p^s$ be an odd prime power such that $p\equiv
1 \mod 4$, $s$ any integer or $p \equiv 3 \mod 4$ and $s$ even, and
$n=2mp^r$ be an oddly even integer with $(m,p)=1$. Then there exists
a negacyclic self-dual code of length $2mp^s$ over $\mathbb{F}_{p^s}$ if and
only if $\ord_m(p^s)$ is odd.
\end{thm}
\pf  Under the hypothesis on $p$, $s$ and $m$ we have from
Theorem~\ref{thm:prod} that the polynomial $x^{2mp^r}+1= \prod
f_i(\gamma x)^{p^r}(x) \prod f_j^{p^r}(-\gamma x) $, where $f_i(x)$
and $f_j(x)$ are the monic irreducible factors of $x^m-1$ in
$\mathbb{F}_{p^s}$. By Lemma~\ref{lem:rev2}, $\ord_m(p^s)$ is odd if and
only if there is no cyclotomic class such that $Cl_m(i)=Cl_m(-i)$.
From Lemma~\ref{lem:rev1}, this is equivalent to saying that there
are no irreducible factors of $x^m-1$ such that $f_i(x)=f_i^*(x)$.
From the ring isomorphisms given in Lemma~\ref{lem:equiv}, we have
that $f_i(x)\ne f_i^*(x)$ for all $i$ is true if and only if
$f_{i}(\gamma x)\neq f^*_i(\gamma x)$ and $f_{j}(-\gamma x)\neq
f^*_{j}(-\gamma x)$ are true. Then from Theorem~\ref{th:exist}
self-dual negacyclic codes exist. \qed
\begin{ex}
A self-dual negacyclic code of length $70$ over $\mathbb{F}_5$ does not
exist. There is no self-dual negacyclic code of length 30 over
$\mathbb{F}_9$, but there is a self-dual code over $\mathbb{F}_9$ of length $126$.
\end{ex}
\begin{lem}
\label{lem:quadra1} Let $p$ and $q$ be distinct odd primes such that
$p$ is not a quadratic residue modulo $q$. Then we have the following.
\begin{enumerate}
\item[(i)] If $q \equiv 1 \pmod 4$, then $\ord_q(p) \equiv 0 \pmod 4$.
\item[(ii)] If $q\equiv 3 \pmod 4$, then $\ord_q (p) \equiv 0 \pmod 2$.
\end{enumerate}
\end{lem}
\pf Assume that $p$ is not a quadratic residue modulo $q$. Then
from~\cite[Lemma 6.2.2]{huffman03} $\ord_q (p)$ is not a divisor of
$\frac{p-1}{2}$, so from Fermat's Little Theorem $\ord_q (p)=q-1$.
Hence $\ord_q(p)\equiv 0 \pmod 4$ since $q \equiv 1 \pmod 4$. If
$q\equiv 3 \pmod 4$, then $\ord_q (p)=q-1 \equiv 0 \pmod 2$. \qed
\begin{lem}
\label{lem:quadra2} Let $n$ be a positive integer and $q$ a prime
power such that $(q,n)=1$. Then we have the following.
\begin{enumerate}
\item[(i)] If $\ord_n(q)$ is even, then $\ord_n(q^2)=\frac{\ord_n(q)}{2}.$
\item[(ii)] If $\ord_n(q)$ is odd, then $\ord_n(q^2)=\ord_n(q).$
\end{enumerate}
\end{lem}
\pf Let $r=\ord_n(q)$ and $r'=\ord_n(q^2).$ Then we have
$q^{2r'}\equiv 1 \mod n,$ which implies that $r|2r'$. Since $r$ is
even, we have $(q^2)^{\frac{r}{2}}=q^r \equiv 1 \mod n$, and then
$r'|\frac{r}{2}$. Hence we obtain that $r'=\frac{r}{2}$. This proves
part (i). For part (ii), assume again that $r=\ord_n(q)$ is odd and
$r'=\ord_n(q^2)$. We then have that $r|2r'$, and since $r$ is odd it
must be that $r|r'$. On the other hand, we have $q^{2r} \equiv n$,
so that $r'|r$, and therefore $r=r'$. \qed

\begin{cor}
\label{cor:quadra3} Let  $p$ and $q$ be two distinct primes such
that $p$ is not a quadratic residue modulo $q$. Then if $q \equiv 1
\pmod 4$ and $p \equiv 1 \pmod 4$, there is no self-dual negacyclic
code of length $2pq^{\alpha}$ over $\mathbb{F}_p$ or $\mathbb{F}_{p^2}$.
\end{cor}
\pf From Lemma~\ref{lem:quadra1}, if $q \equiv 1 \pmod 4$ and $p$ is
not a quadratic residue modulo $q$, then $\ord_q(p) \equiv 0 \pmod
4$. Hence from Lemma~\ref{lem:quadra2}, $\ord_q(p^2)$ is even. Then
the proof of Lemma~\ref{lem:rev2} implies that
$\ord_{q^\alpha}(p^2)$ is even. Hence from~Theorem~\ref{th:self}
there are no self-dual negacyclic codes of length $2q^{\alpha}p$
over $\mathbb{F}_{p}$ or $\mathbb{F}_{p^2}.$ \qed
\begin{ex}
For $p=5$, $q=13$ and $q=17$ satisfy the hypothesis of
Corollary~\ref{cor:quadra3}. Hence there are no self-dual negacyclic
codes over $\mathbb{F}_5$ and $\mathbb{F}_{25}$ with lengths 130, 170 or 1690.
\end{ex}

\section{Repeated Root Cyclic Codes}

It is well known that the cyclic codes of length $n$ over $\mathbb{F}_{p^s}$
are principal ideals of $\mathbb{F}_{p^s}[x]/(x^n-1)$, and these ideals are
generated by the monic factors of $x^n-1$. Hence the importance of
the decomposition of the polynomial $x^n-1$ over $\mathbb{F}_{p^s}$.

Let $n=2mp^r$, with $m$ an odd integer such that $(m,p)=1$. Then we
have the decomposition
$x^n-1=(x^{2m}-1)^{p^r}=(x^{m}-1)^{p^s}(x^{m}+1)^{p^r}$. Since
$(m,p)=1$, the polynomials $x^{m}-1$ and $x^{m}+1$ factor uniquely
as the product of monic irreducible pairwise coprime polynomials
given by $x^{m}-1= \prod_{i =1}^{k} f_i$ and $x^{m}+1=
\prod_{j=1}^{l}g_j$. This is due to the fact that $(m,p)=1$, so the
roots are simple~\cite[Proposition 2.7]{permounth}. Let $f_i(x)$ be
a monic irreducible divisor of $x^m-1$. Then there exists $h(x) \in
\mathbb{F}_{p^s}[x]$ such that $f_i(x)h(x)=x^m-1$, and hence
$f_i(-x)h(-x)=(-x)^m-1=-(x^m+1)$. Therefore $f_i(-x)$ is a monic
irreducible divisor of $x^m+1$. This gives that the factorization of $x^n-1$ is
\[
x^{2mp^r}-1=\prod_{i =1}^{k} (f_i(x)f_i(-x))^{p^r}.
\]
Hence a cyclic code of length $n=2mp^r$ over $\mathbb{F}_{p^s}$ is of the form
\[
C= \left\langle  \prod (f_i(x))^{\alpha_i}  \prod
(f_j(-x))^{\beta_j} ) \right\rangle,
\]
where $0 \le \alpha_i, \beta_j  \le p^r,$ $ 1 \le i,j \le k$,
and $f_i$, $i \le k$ is an irreducible factor of $x^m-1$.
This gives the following result.
\begin{prop}
For $p$ an odd prime, the cyclic codes of length $n=2mp^r$, $m$ an
odd integer such that $(m,p)=1$, are generated by
\[
\left\langle  \prod (f_i(x))^{\alpha_i}  \prod (f_j(-x))^{\beta_j} \right\rangle,
\]
where $0 \le \alpha_i, \beta_j  \le p^r,$ $ 1 \le i,j \le k$, and $f_i$, $i \le k$,
is a monic irreducible factor of $x^m-1$.
\end{prop}

\subsection{The Number of Cyclic Self-dual Codes}

It has been proven~\cite{dinh,jia,xai} that cyclic self-dual codes
exist if and only if the characteristic is 2. Since a self-dual
cyclic code must have even length and characteristic 2, cyclic
self-dual codes have repeated roots. In \cite[Corollary 2]{jia}, Jia
et al. gave the number of self-dual cyclic codes in some cases. The
remainder of this characterization was left as an open problem,
namely the case when the length of the code contains a prime factor
congruent to $1 \mod 8$. The following proposition is used in
answering this problem.
\begin{prop}
\label{prop:final} Let $p\equiv 1 \mod 8$ be an odd prime number,
and $m$ be an odd number. Then we have the following implication
\[\ord_p(2)=2^ke \Rightarrow \forall\, 0 \le l\le k, \, \ord_p(2^{2^l})=2^{k-l}e .\]
\end{prop}
\pf Since $p\equiv 1 \mod 8$, from~\cite[Lemma 6.2.5]{huffman03} $2$
is a quadratic residue modulo $p$. Hence $\ord_p(2) |
\frac{p-1}{2}$, i.e.,  $\ord_p(2)=2^ke$ for some $k>0$. Then
from~Lemma~\ref{lem:quadra2} (i) we have that
$\ord_p(2^2)=2^{k-1}e$. Using the same argument $l$ times, the
result follows. \qed

\begin{cor}
Let $n=2^{r}p^\alpha$. Then there is a unique cyclic self-dual code
of length $n$ over $\mathbb{F}_{2^s}$ generated by
$g(x)=(x^{p^\alpha}+1)^{2^{r-1}}$ in the following cases
\begin{enumerate}
\item[(i)] $p \equiv 3 \mod 8, s \text{ odd}$,
\item[(ii)] $p \equiv 5 \mod 8, s \text{ odd or } s \equiv 2 \mod 4$,
\item[(iii)] $p \equiv 1 \mod 8$ and $\ord_p(2)=2^ke$, and $s={2^l}$,\\
\hspace*{0.3in}$0<l<k$.
\end{enumerate}
\end{cor}
\pf Parts (i) and (ii) follow from~\cite[Proposition 2]{jia}.
When $p \equiv 1 \mod 8$ and $\ord_p(2)=2^ke$, for $s=2^l$ with $0<l <k$,
from Proposition~\ref{prop:final} we have that $\ord_p(2^s)$ is an
even integer.
Hence from~\cite[Theorem 4]{jia} there is a unique
self-dual code generated by $g(x)=(x^{p^\alpha}+1)^{2^{r-1}}$.
\qed

\begin{ex}
Let $r$ and $\alpha$ be positive integers.
\begin{enumerate}
  \item [(i)] For $p=3$ and $s = 5$, the polynomial $g(x) = (x^{3^{\alpha}}+1)^{2^{r-1}}$
  generates the unique self-dual cyclic code of length $2^{r}3^{\alpha}$ over $\mathbb{F}_{2^{5}}$.
  \item [(ii)] For $p=5$ and $s = 6$, the polynomial $g(x) = (x^{5^{\alpha}}+1)^{2^{r-1}}$
  generates the unique self-dual cyclic code of length $2^{r}5^{\alpha}$ over $\mathbb{F}_{2^{6}}$.
  \item [(iii)] For $p=17$, $\ord_{17}(2)=2^3$, so $l=2$. Then $g(x) = (x^{17^{\alpha}}+1)^{2^{r-1}}$
  generates the unique self-dual cyclic code of length $2^{r}17^{\alpha}$ over $\mathbb{F}_{2^{2}}$.
\end{enumerate}
\end{ex}
\balance


\begin{thebibliography}{99}

\bibitem{aicha2011}{A. Batoul, K. Guenda, and T. A. Gulliver},
``On self-dual cyclic codes over finite chain rings,''
{\em Design. Codes Crypt.}, to appear.

\bibitem{berl1}{E. R. Berlekamp},
``Negacyclic codes for the Lee metric,''
in {\em Proc. Conf. Combinatorial Mathematics and its Applic.},
Univ. North Carolina Press, Chapel Hill, NC, pp. 298--316, 1968.

\bibitem{blackford}{T. Blackford},
``Negacyclic duadic codes,''
{\em Finite Fields Appl.}, vol. 14, no. 4, pp. 930--943, Nov. 2008.

\bibitem{castagnoli}{G. Castagnoli, J. L. Massey, P. A. Schoeller, and N. von Seemann},
``On repeated-root cyclic codes,''
{\em IEEE Trans. Inform. Theory}, vol. 37, no. 2, pp. 337--342, Mar. 1991.

\bibitem{demazure}{ M. Demazure,}
{\em Cours D'Alg\`{e}bre: Primalit\'{e}, Divisibilit\'{e}, Codes},
Cassini, Paris, 1997.

\bibitem{dinh}{H. Q. Dinh},
``Repeated-root constacyclic codes of length $2p^s$,''
{\em Finite Fields Appl.}, vol. 18, no. 1, pp. 133--143, Jan. 2012.

%
\bibitem{permounth} H. Dinh and S. R. L\'opez-Permouth,
``Cyclic and negacyclic codes over finite chain rings,''
{\em IEEE Trans. Inform. Theory}, vol. 50, no. 8, pp. 1728--1744, Aug. 2004.

\bibitem{guenda011}{K. Guenda},
``New MDS self-dual codes over finite fields,''
{\em  Des. Codes Crypt.}, vol. 62, no. 11, pp. 31--42, Jan. 2012.

\bibitem{huffman03}{W. C. Huffman and V. Pless},
{\em Fundamentals of Error-Correcting Codes}, Cambridge Univ. Press, New York, 2003.


\bibitem{xai}{X. Kai and S. Zhu},
``On cyclic self-dual codes,''
{\em Appl. Algebra Engr. Commun. Comput.}, vol. 19, pp. 509--525, 2008.

\bibitem{jia}{Y. Jia, S. Ling, and C. Xing},
``On self-dual cyclic codes over finite fields,''
{\em IEEE Trans. Inform. Theory}, vol. 57, no. 4, pp. 2243--2251, Apr. 2011.
%

\bibitem{vanlint}{J. H. van Lint},
``Repeated-root cyclic codes,''
{\em IEEE Trans. Inform. Theory},
vol. 37, no. 2, pp. 343--345, Mar. 1995.

\end{thebibliography}
\end{document}